\documentclass[twocolumn,prb,showpacs,floats]{revtex4}
\usepackage[dvips]{graphicx}
\usepackage{dcolumn}
\usepackage{color}

\usepackage{multirow}

\usepackage{bm}

\newcommand{\SNIO}{Sr$_2$NiIrO$_6$}
\newcommand{\SZIO}{Sr$_2$ZnIrO$_6$}

\begin{document}

\title{Long-range magnetic interaction and frustration in double perovskite Sr$_2$NiIrO$_6$}
\author{Xuedong Ou, Zhengwei Li, Fengren Fan, Hongbo Wang,}
\affiliation{Laboratory for Computational Physical Sciences (MOE), State Key Laboratory of Surface Physics, and Department of Physics, Fudan University, Shanghai 200433, China}
\author{Hua Wu}
\thanks{Corresponding author: wuh@fudan.edu.cn}
\affiliation{Laboratory for Computational Physical Sciences (MOE), State Key Laboratory of Surface Physics, and Department of Physics, Fudan University, Shanghai 200433, China}

\date{today}

\begin{abstract}

Sr$_2$NiIrO$_6$ would be a ferromagnetic (FM) insulator in terms of the common superexchange mechanism between the first nearest neighboring (1NN) magnetic ions Ni$^{2+}$ ($t_{2g}^6e_g^2$) and Ir$^{6+}$ ($t_{2g}^3$). However, the observed antiferromagnetic (AF) order questions this viewpoint. In this work, we present first-principles calculations and find that while the 1NN Ni$^{2+}$-Ir$^{6+}$ exchange is indeed FM, the 2NN and 3NN couplings in the fcc Ir (and Ni) sublattice are AF. Moreover, the 2NN AF Ir-Ir coupling turns out to be even stronger than the 1NN FM Ni-Ir coupling, thus giving rise to a magnetic frustration. Sr$_2$NiIrO$_6$ hence becomes a distorted low-temperature antiferromagnet. Naturally, a very similar magnetic property in Sr$_2$ZnIrO$_6$ can be explained by the frustrated AF coupling in the fcc Ir$^{6+}$ sublattice. This work highlights the long-range magnetic interaction of the delocalized $5d$ electrons, and also addresses why the spin-orbit coupling is ineffective here.

\end{abstract}

\pacs{75.25.Dk, 71.20.-b, 71.70.-d}

\maketitle
\section{Introduction}

In the insulating transition-metal (TM) oxides, superexchange (SE) coupling of neighboring magnetic ions via intermediate oxygen, according to the Goodenough-Kanamori-Anderson rules~\cite{GKA}, commonly plays a leading role in their magnetic order. One simple but useful rule is that for a linear $M$-O-$M'$ exchange path, the SE would be antiferromagnetic (AF) [ferromagnetic (FM)] when the active orbitals of $M$ and $M'$ are same [different]. Fig. 1(a) shows two $d^1$ ions each having two orthogonal A-B levels and the same A-level occupation. Taking into account an effective hopping $t$ between
two ions associated with the charge fluctuation ($d^1 + d^1 \rightarrow d^0 + d^2$) where the electron correlation Hubbard $U$ is involved, an energy gain
of an AF order (relative to a FM one) is proportional to $t^2$/$U$ in a strong correlation limit ($U\gg~t$). Fig. 1(b) shows two different $d^1$ level occupations, and a FM stability against AF is proportional to $t^2J_{\rm H}$/$U^2$ where $J_{\rm H}$ is a Hund exchange. This is the reason why a FM Mott insulator is often associated with orbital physics (e.g., an orbital ordering) and its $T_{\rm C}$ is much lower (due to the factor $J_{\rm H}$/$U$ $\sim$ 1/5) than the $T_{\rm N}$ of many AF Mott insulators.

In practice, it is often sufficient to consider the SE between the nearest neighboring (NN) magnetic ions only. This approach applies with much success to numerous insulating $3d$ TM oxides,
where the $3d$ electrons are quite localized due to the strong correlation effect. In recent years, $5d$ TM oxides have received considerable attention due to their significant spin-orbit coupling (SOC) effect and possibly exotic properties~\cite{Kim08,Kim09,Jackeli,Wan,Mazin,Yin,Ou1,Ou2,Cao}. The hybrid $3d$-$5d$ TM oxides are also of current great interest for exploration of novel magnetic and electronic properties in this material system, in which new SOC effects
add to the common charge-spin-orbital physics appearing in the $3d$ TM oxides. Among them, the double perovskites $A_2BB'$O$_6$ ($A$ = alkaline earth metal, $B$ = $3d$ TM, and $B'$ = $5d$ TM) are an important material platform~\cite{Tokura99,Serrate,Alff,Meetei,Paul,Morrow,Kayser,Yan,Feng,
Morrow2,Wang,Kanungo}: Sr$_2$FeReO$_6$ is an above room temperature (RT) ferrimagnetic half metal~\cite{Tokura99}, and Sr$_2$CrOsO$_4$ is a ferrimagnetic insulator with a seemingly highest $T_{\rm C}$ in the perovskite oxides~\cite{Alff,Meetei}, etc. As $5d$ electrons are moderately or weakly correlated and their orbitals are much delocalized, their magnetic coupling could well be a long-range interaction.

\begin{figure}[b]
\centering \includegraphics[width=7cm]{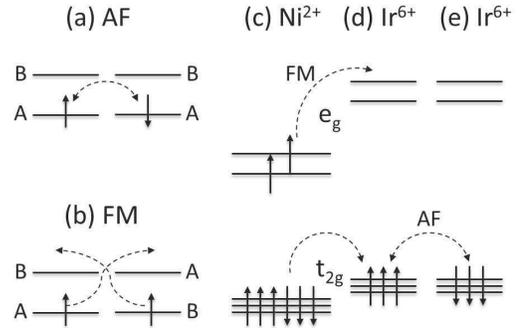}
\caption{
(a) AF and (b) FM SE between two two-level $d^1$ ions.
(c) and (d): {\SNIO} would be FM, according to the SE between the NN Ni$^{2+}$ and Ir$^{6+}$ ions. (d) and (e): AF SE in the fcc Ir$^{6+}$ sublattice.
}
\end{figure}

In this work, we study the electronic structure and magnetism of the newly synthesized double perovskite {\SNIO},~\cite{Kayser} using density functional calculations. This material crystallizes in the monoclinic space group $P2_1/n$ at RT (see Fig. 2) and undergoes two structural phase transitions ($P2_1/n$ $\rightarrow$ $I4/m$ $\rightarrow$ $Fm$\=3$m$) upon heating. Magnetic susceptibility measurements~\cite{Kayser} suggest the establishment of AF interactions at $T_{\rm N}$ = 58 K. This oxide has the Ni$^{2+}$ ($t_{2g}^6e_g^2$)-Ir$^{6+}$ ($t_{2g}^3$) charge state as seen below. Taking into account a charge fluctuation into the common Ni$^{3+}$-Ir$^{5+}$ state (a reverse Ni$^+$-Ir$^{7+}$ is quite unusual), both the Ni up-spin $e_g$ and down-spin $t_{2g}$ electron hopping (the Ni up-spin $t_{2g}$ levels lie lowest due to the crystal field splitting and Hund exchange) would give a FM SE between the Ni$^{2+}$ and Ir$^{6+}$ ions, see Figs. 1(c) and 1(d). As the $e_g$ and $t_{2g}$ levels are orthogonal, the $e_g$ ($t_{2g}$) electron hopping follows the simple SE mechanism plotted in Fig. 1(b) [Fig. 1(a)]. Apparently, this expected FM order contradicts the observed AF in {\SNIO}, and thus consideration of only NN Ni$^{2+}$-Ir$^{6+}$ coupling would be a mistake here. Then, a possibly long-ranged Ir-Ir coupling within the fcc sublattice should be invoked, which would be AF due to the half-filled $t_{2g}^3$ shells [Figs. 1(d) and 1(e)]. As we calculate below, there is indeed a long-range AF interaction in the fcc Ir$^{6+}$ sublattice, and the second NN Ir-Ir AF coupling energy is even bigger than the first NN Ni-Ir FM one, thus giving rise to a magnetic frustration. As a result, {\SNIO} behaves as a distorted low-temperature antiferromagnet~\cite{Kayser}. Naturally, the frustrated AF couplings in the fcc Ir$^{6+}$ sublattice explain a very similar magnetic property in the isostructural {\SZIO}.~\cite{Kayser} Note that one could take care of long-range magnetic interaction of the delocalized $5d$ electrons.

\begin{figure}[t]
\centering \includegraphics[width=6cm]{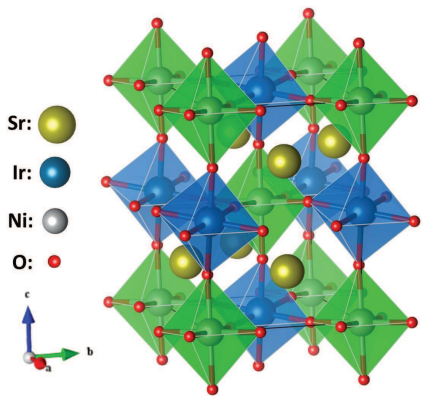}
\caption{(Color online)
Double perovskite structure of {\SNIO}. The Ni and Ir ions form their respective fcc sublattices.
}
\end{figure}

\section{Computational details}

Our calculations were performed using the full-potential augmented plane waves plus local orbital method (WIEN2K code)~\cite{Blaha}. We took the structure data of {\SNIO} measured by neutron diffraction at RT~\cite{Kayser}. The muffin-tin sphere radii are chosen to be 2.8, 2.1, and 1.5 Bohr for Sr, Ni/Ir, and O atoms, respectively. The cutoff energy of 16 Ry is used for plane wave expansion of interstitial wave functions, and 6$\times$6$\times$4 {\bf k} mesh for integration over the Brillouin zone, both of which ensure a sufficient numerical accuracy. SOC is included by the second-variational method with scalar relativistic wave functions. We employ the local spin density approximation plus Hubbard $U$ (LSDA+$U$) method~\cite{Anisimov} and use the typical values, $U$ = 6 eV and $J_{\rm H}$ = 0.9 eV ($U$ = 2 eV and $J_{\rm H}$ = 0.4 eV), to describe electron correlation of the Ni $3d$ (Ir $5d$) electrons.
The calculated Mott insulating state of {\SNIO} remains unchanged in a reasonable range of the $U$ values ($U$ = 4-8 eV for Ni $3d$ and $U$ = 1-3 eV for Ir $5d$), and the corresponding variation of 1-2 meV for the exchange energy parameters does not affect our discussion and conclusion about the frustrated magnetism.

\section{Results and discussion}

\begin{figure}[t]
\centering \includegraphics[width=7cm]{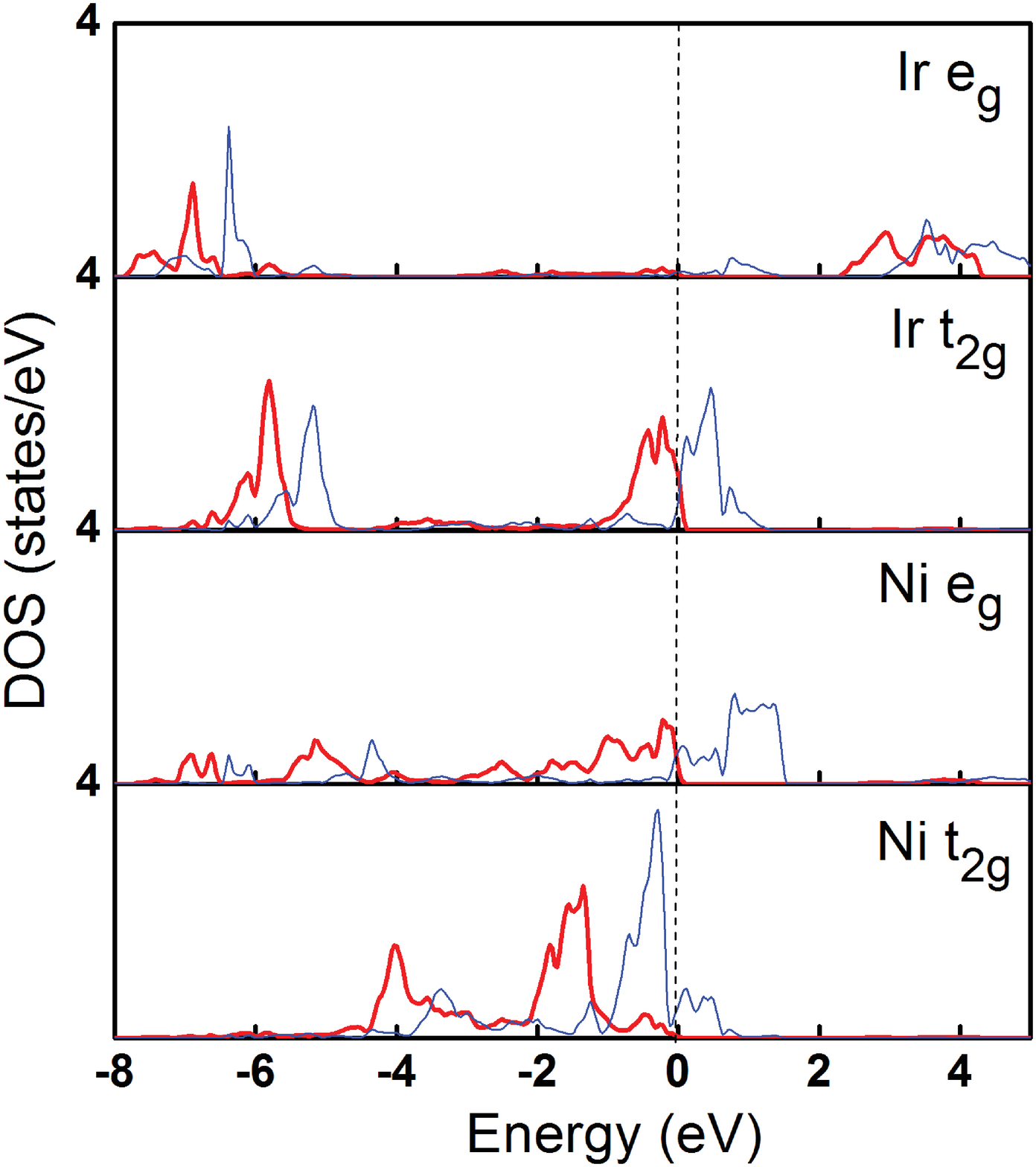}
\caption{(Color online)
Ir $5d$ and Ni $3d$ DOS of {\SNIO} calculated by LSDA for the FM state. The solid red (thin blue) curves stand for the up (down) spin channel. Fermi level is set at zero energy. {\SNIO} has the Ni$^{2+}$ ($t_{2g}^6e_g^2$)-Ir$^{6+}$ ($t_{2g}^3$) charge state.
}
\end{figure}

We first study the electronic structure of {\SNIO} and the Ni-Ir charge state. Fig. 3 shows the orbitally resolved density of states (DOS) calculated by LSDA for the FM state. The delocalized Ir $5d$ electrons have a strong covalency with the ligand oxygens, giving rise to a large bonding-antibonding splitting. The $pd\sigma$ splitting of the Ir $e_g$ electrons is up to 9 eV,
and the $pd\pi$ splitting of the Ir $t_{2g}$ electrons is about 6 eV.
The Ir $5d$ electrons have a $t_{2g}$-$e_g$ crystal-field splitting of more than 3 eV. Besides the occupied bonding states (around --6 eV) ascribed to the lower-lying O $2p$ bands, only the up-spin Ir $t_{2g}$ state is occupied, giving a formal Ir$^{6+}$ charge state with a $t_{2g}^3$ ($S$ = 3/2) configuration. In contrast, the Ni $3d$ electrons are confined and have a smaller $pd\sigma$ ($pd\pi$) bonding-antibonding splitting of 4 eV (2 eV) and the $t_{2g}$-$e_g$ crystal-field splitting of 1-1.5 eV. Only the down-spin Ni $e_g$ antibonding state is unoccupied, giving a formal Ni$^{2+}$ charge state with the $t_{2g}^6e_g^2$ ($S$ = 1) configuration. Therefore, {\SNIO} has the Ni$^{2+}$-Ir$^{6+}$ charge state. Its closed subshells and a finite electron correlation would certainly make {\SNIO} insulating. However, in the present LSDA calculation, the bandwidth of the Ir $t_{2g}$ electrons is slightly larger than the exchange splitting, making the Ir $t_{2g}$ bands of two spin directions somewhat overlapping at the Fermi level. As seen below, this metallic solution will turn into a Mott insulating one upon inclusion of the electron correlation.

\begin{figure}[t]
\centering \includegraphics[width=7cm]{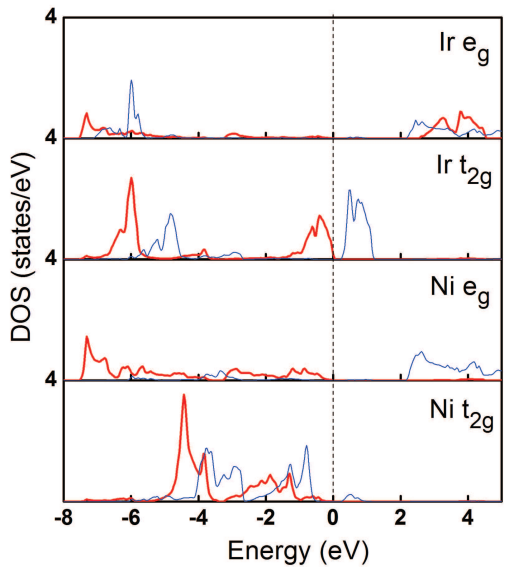}
\caption{(Color online)
Insulating band structure of {\SNIO} in the Ni$^{2+}$ ($t_{2g}^6e_g^2$)-Ir$^{6+}$ ($t_{2g}^3$) charge state calculated by LSDA+$U$ for the FM state. Other magnetic states have a very similar band structure.
}
\end{figure}

We now include the static electron correlation by carrying out LSDA+$U$ calculations. The insulating band structure is shown in Fig. 4. It has a small band gap of 0.3 eV within the Ir $t_{2g}$ bands due to the moderate electron correlation of the delocalized Ir $5d$ electrons. The Ni $3d$ bands have a gap of more than 2 eV due to the strong correlation. The electron correlation enhances electron localization and reduces band hybridization and further stabilizes the Ni$^{2+}$-Ir$^{6+}$ charge state~\cite{footnote}. The Ni$^{2+}$ ($S$ = 1) ion has a spin moment of 1.76 $\mu_B$ (see Table I), being close to its formal value of 2 $\mu_B$. The Ir$^{6+}$ ($S$ = 3/2) ion has a smaller moment of 1.46 $\mu_B$ reduced by the strong covalency with the oxygen ligands.

\begin{table}[b]
\caption{Relative total energies $\Delta{E}$ (meV/fu) and spin moments $M$ (in unit of $\mu_B$) calculated by LSDA+$U$ for different systems in different magnetic states. The Ir-Ir magnetic interactions are estimated for Sr$_2$ZnIrO$_6$ either in {\SNIO} structure (Zn substitution for Ni) or in its real structure~\cite{Kayser}. The Ni-Ni exchange coupling is estimated using the artificial La$_2$NiSiO$_6$ in {\SNIO} structure. The derived exchange energy parameters (meV) for the 1NN Ni-Ir, 2NN Ir-Ir and Ni-Ni, and 3NN Ir-Ir pairs are listed in the last two lines.}
\begin{tabular}{l@{\hskip3mm}c@{\hskip3mm}c@{\hskip3mm}c}
\hline\hline
 System & Magn. & $\Delta{E}$ & $M$ (Ni$^{2+}$/Ir$^{6+}$) \\ \hline
{\SNIO} & FM & 0 & 1.76/1.46  \\
        & G-AF & 89 & 1.64/1.28  \\ \hline
Sr$_2$Zn(Ni)IrO$_6$ & FM & 0 & /1.39  \\
        &  layered AF & --84 & /1.31  \\ \hline
La$_2$NiSiO$_6$ & FM & 0 & 1.70/  \\
        &  layered AF & --19 & 1.69/  \\ \hline
Sr$_2$ZnIrO$_6$ & FM & 0 & /1.42  \\
        &  layered AF & --75 & /1.34  \\
        &  bilayered AF & --42 & /1.36  \\  \hline
 $J_{\rm Ni-Ir}$ & $J'_{\rm Ir-Ir}$ & $J'_{\rm Ni-Ni}$ & $J''_{\rm Ir-Ir}$ \\
 --7.4 & 9.4, 10.5 & 2.4 & 2.2 \\
\hline\hline
\end{tabular}
\end{table}

As both the Ni$^{2+}$ and Ir$^{6+}$ ions are magnetic and form their respective fcc sublattices, their magnetic interactions are of concern. Here we study different magnetic structures using LSDA+$U$ calculations. The G-AF state of {\SNIO} (FM Ni$^{2+}$ and Ir$^{6+}$ sublattices being AF coupled) turns out to be less stable than the FM state by 89 meV/fu, see Table I. As the FM and G-AF states differ in the exchange energy only by the 1NN Ni-Ir couplings, which are $\pm$6$J_{\rm Ni-Ir}$ per formula unit. Then the average exchange energy parameter of the 1NN Ni-Ir pairs can be estimated to be $J_{\rm Ni-Ir}$ = --89/12 $\approx$ --7.4 meV. This FM Ni-Ir coupling is readily understood by a SE mechanism, see Fig. 1 and the Introduction. However, the observed AF interaction~\cite{Kayser} at $T_{\rm N}$ = 58 K questions this description. Therefore, we are motivated to study the long-range magnetic interactions, particularly associated with the delocalized Ir $5d$ electrons. To do so, we use two artificial systems with either Ir$^{6+}$ or Ni$^{2+}$ magnetic sublattice only, Sr$_2$ZnIrO$_6$ [i.e., Sr$_2$Zn(Ni)IrO$_6$ in Table I] and La$_2$NiSiO$_6$ both in the {\SNIO} structure, to calculate the 2NN Ir$^{6+}$-Ir$^{6+}$ and Ni$^{2+}$-Ni$^{2+}$ exchange parameters ($J'_{\rm Ir-Ir}$ and $J'_{\rm Ni-Ni}$ with a reference to the 1NN $J_{\rm Ni-Ir}$). This approach avoids choices of complicate magnetic structures in bigger supercells, and allows to estimate the two parameters separately.
For Sr$_2$Zn(Ni)IrO$_6$, the layered AF state (FM $ab$ planes being AF alternate along the $c$ axis, see also Fig. 2) is more stable than the FM state by 84 meV/fu, see Table I. The layered AF and FM states differ in the exchange energy only by the 2NN Ir-Ir couplings (with a reference to the 1NN Ni-Ir ones), i.e., --2$J'_{\rm Ir-Ir}$ $vs$ 6$J'_{\rm Ir-Ir}$. Then the energy difference gives AF $J'_{\rm Ir-Ir}$ = 84/8 = 10.5 meV. The corresponding energy difference of 19 meV/fu for La$_2$NiSiO$_6$ gives AF $J'_{\rm Ni-Ni}$ = 19/8 $\approx$ 2.4 meV, see Table I.

As the magnetic Ir$^{6+}$ and Ni$^{2+}$ ions have closed subshells, the SE interactions naturally explain the AF $J'_{\rm Ir-Ir}$ and $J'_{\rm Ni-Ni}$. Note that the Ni$^{2+}$ $3d$ electrons are confined but the Ir$^{6+}$ $5d$ electrons are delocalized, it is therefore not surprising that $J'_{\rm Ir-Ir}$ is about four times as big as $J'_{\rm Ni-Ni}$. However, it is a bit surprising that the 2NN AF $J'_{\rm Ir-Ir}$ is even bigger than the 1NN FM $J_{\rm Ni-Ir}$, thus giving rise to a magnetic frustration in {\SNIO}.
This vital role of the strong 2NN AF Ir-Ir coupling is also manifested in the real double perovskite Sr$_2$ZnIrO$_6$, see below.

Sr$_2$ZnIrO$_6$ has a very similar crystal structure and magnetic property to {\SNIO}, and it has AF interactions at $T_{\rm N}$ = 46 K.~\cite{Kayser} We have also calculated different magnetic states of Sr$_2$ZnIrO$_6$ and find the 2NN AF $J'_{\rm Ir-Ir}$ = 75/8 $\approx$ 9.4 meV (see Table I), being close to $J'_{\rm Ir-Ir}$ = 10.5 meV in {\SNIO}.
As the delocalized Ir $5d$ electrons produce a long-range magnetic interaction, we also estimate the 3NN AF $J''_{\rm Ir-Ir}$ (the exchange path along the linear Ir-O-Ni-O-Ir bonds with the Ir-Ir distance of 7.8 \AA) by calculating the bilayered AF state of Sr$_2$ZnIrO$_6$. The bilayered AF state has FM $ab$ planes but AF alternation every bilayer along the $c$ axis, and it is more stable than the FM state by 42 meV/fu. The exchange energy per formula unit can be expressed as 6$J'_{\rm Ir-Ir}$ + 3$J''_{\rm Ir-Ir}$ for the FM state and 2$J'_{\rm Ir-Ir}$ + $J''_{\rm Ir-Ir}$ for the bilayered AF state. Therefore, the AF $J''_{\rm Ir-Ir}$ is estimated to be (42 -- 4 $\times$ 9.4 ) / 2 = 2.2 meV.

As seen from the above results, apparently the Ir-Ir magnetic interactions are long-ranged and have a non-negligible strength even at a distance of about 8 \AA. It is the long-range AF interactions of the Ir$^{6+}$ sublattice which make Sr$_2$ZnIrO$_6$ magnetically frustrated. It is the strongest 2NN AF $J'_{\rm Ir-Ir}$ which overwhelms the 1NN FM $J_{\rm Ni-Ir}$ and also makes {\SNIO} magnetically frustrated. In a word, the long-range magnetic interactions and frustration would make the cubic double perovskites {\SNIO} and Sr$_2$ZnIrO$_6$ distorted, and this would partially relieve the magnetic frustration and eventually stabilize them into a very similar low-temperature antiferromagnet as experimentally observed~\cite{Kayser}.

\begin{figure}[t]
\centering \includegraphics[width=7cm]{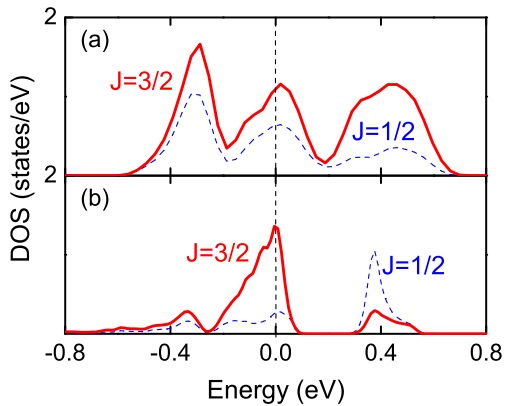}
\caption{(Color online)
The LDA+SOC calculated Ir$^{6+}$ $t_{2g}$ DOS projected onto the SOC basis set, the $J$ = 3/2 quartet (solid red curves) and the $J$ = 1/2 doublet (dashed blue curves). (a) In Sr$_2$ZnIrO$_6$, the overall mixing of the $J$ = 3/2 and $J$ = 1/2 states is due to the band formation of the delocalized Ir $5d$ electrons in the fcc Ir sublattice with twelve Ir-Ir coordination. (b) The SOC splitting of about 0.5 eV between the $J$ = 3/2 and the $J$ = 1/2 states is restored upon the reduction of the Ir-Ir coordination to four, which is modeled in the artificial system Sr$_2$GaIr$_{0.5}$Si$_{0.5}$O$_6$ (in Sr$_2$ZnIrO$_6$ structure) with alternating GaIr and SiGa planes.
}
\end{figure}

Finally, we check if the SOC is important or not in the present materials. Normally, SOC is important in heavy $5d$ TMs, and particularly, iridates recently receive great interest~\cite{Kim08,Kim09,Jackeli,Wan,Mazin,Yin,Ou1,Ou2,Cao}. Owing to a large crystal-field splitting, iridates are in a low-spin state with only the $t_{2g}$ occupation (e.g., in a cubic crystal field). Then the SOC splits the $t_{2g}$ triplet (with 2-fold spin degeneracy) into the lower $J$ = 3/2 quartet and the higher $J$ = 1/2 doublet~\cite{Kim08,Kim09}. We have used this SOC basis set to project the Ir$^{6+}$ $t_{2g}$ DOS of Sr$_2$ZnIrO$_6$ calculated by LDA+SOC, but we find that the $J$ = 3/2 and the $J$ = 1/2 states are completely mixed, see Fig. 5(a). Therefore, the $J$ = 3/2 and the $J$ = 1/2 states are not at all eigen orbitals in Sr$_2$ZnIrO$_6$ (and in {\SNIO} with the same fcc Ir$^{6+}$ sublattice). This is because the delocalized Ir $5d$ electrons form, with the intersite electron hoppings in the fcc sublattice (the high coordination of twelve), a `broad' band with its bandwidth being more than 1 eV. Then the SOC effect is `killed'. In contrast, if the Ir-Ir coordination number is reduced as in the low-dimensional iridates, the SOC effect would be manifested. To check this, we also calculate the artificial system Sr$_2$GaIr$_{0.5}$Si$_{0.5}$O$_6$ (in Sr$_2$ZnIrO$_6$ structure) with alternating GaIr and SiGa planes. The Ga$^{3+}$, Ir$^{6+}$ and Si$^{4+}$ ions have well comparable ionic sizes, and they make charge balanced and the Ir$^{6+}$-Ir$^{6+}$ ions only four-coordinated. In this case, the SOC splitting of about 0.5 eV between the $J$ = 3/2 and the $J$ = 1/2 states is well restored as seen in Fig. 5(b), and thus the $J$ = 3/2 and the $J$ = 1/2 states would serve as eigen orbitals in a good approximation~\cite{Ou1}.

The above results show that in Sr$_2$ZnIrO$_6$ and {\SNIO}, the delocalized Ir$^{6+}$ $5d$ electrons have an insignificant SOC effect due to the band formation in the fcc sublattice. Moreover, the half filled $t_{2g}^3$ subshell of the high-valence Ir$^{6+}$ ion has an intrinsic exchange splitting of about 1 eV, see Fig. 3. Both the band effect and the exchange splitting are stronger than the SOC strength, making the SOC ineffective in {\SNIO} and Sr$_2$ZnIrO$_6$. Our LSDA+$U$+SOC test calculations indeed show that the Ir$^{6+}$ ion has only a small orbital moment of 0.07 $\mu_B$, being antiparallel to the spin moment of about 1.3 $\mu_B$ reduced from the formal $S$ = 3/2. Therefore, both Sr$_2$ZnIrO$_6$ and {\SNIO} can be described as an Ir$^{6+}$ $S$ = 3/2 fcc frustrated system, although {\SNIO} itself has an appreciable Ni$^{2+}$-Ir$^{6+}$ FM coupling.

\section{Conclusion}

In summary, using density functional calculations, we find that the newly synthesized isostructural double perovskites {\SNIO} and Sr$_2$ZnIrO$_6$ are insulating and have the formal Ir$^{6+}$ $S$ = 3/2 fcc sublattice, in addition to the Ni$^{2+}$ $S$ = 1 sublattice in the former. The delocalized Ir $5d$ electrons produce long-range magnetic interactions, and the 2NN Ir-Ir AF interaction turns out to be even stronger than the 1NN Ni-Ir FM interaction. Therefore, the leading AF interactions in the fcc Ir sublattice give rise to a magnetic frustration in both {\SNIO} and Sr$_2$ZnIrO$_6$. As a result, both the cubic compounds appear as a distorted low-temperature antiferromagnet. Note that the band formation in the high-coordination fcc Ir sublattice and the exchange splitting of the high-valence Ir$^{6+}$ ion both make the SOC ineffective, and the long-range interactions of the delocalized $5d$ electrons (band formation and magnetic coupling) would be taken care of.   \\

{\bf Acknowledgment.}
This work was supported by the NSF of China (Grant Nos. 11274070 and 11474059), MOE Grant No. 20120071110006, and ShuGuang Program of Shanghai (Grant No. 12SG06).


\begin{thebibliography}{22}
\bibitem{footnote}
We also test the Ni$^{3+}$-Ir$^{5+}$ state, using constrained LSDA+$U$ calculations. We initialize the corresponding occupation number matrix and the orbital polarized potential, and consider a possible $J$ = 0 singlet state of the Ir$^{5+}$ ion due to its strong SOC. After a full electronic relaxation, however, the self-consistent LSDA+$U$+SOC calculations converge also to the Ni$^{2+}$-Ir$^{6+}$ state as reported in the main text.
\end{thebibliography}
\end{document}